\documentclass[fleqn,twoside]{article}
\usepackage{espcrc2}
\usepackage{graphicx}
\usepackage{bm}

\newcommand{\imag}{\mbox{i}\ }

\newcommand{\AmS}{{\protect\the\textfont2
  A\kern-.1667em\lower.5ex\hbox{M}\kern-.125emS}}

\title{Insight into nucleon structure from  generalized
  parton distributions}

\author{LHPC and SESAM Collaborations
{\thanks{Supported by DOE contracts DE-FC02-94ER40818, DE-FG02-91ER40676, and DE-AC05-84ER40150.
 Ph.H. and W. S. are Feodor-Lynen Fellows}}:\\
J.W. Negele\address[MIT]{Center for Theoretical Physics,
        Massachusetts Institute of Technology,
 	Cambridge, MA  02139, USA},
R.C. Brower\address[UBOS]{Department of Physics,
		Boston University,  Boston, MA  02215, USA}, P.
    Dreher\addressmark[MIT],
    R. Edwards\address[JLAB]{Thomas Jefferson National Accelerator Facility,      
    Newport News, VA  23606, USA}, 
    G. Fleming\addressmark[JLAB], Ph.
    H{\"a}gler\addressmark[MIT], 
    Th. Lippert\address[UWUP]{Department of Physics, University of
    Wuppertal, D-42097 Wuppertal, Germany}, \\ A.V.
    Pochinsky\addressmark[MIT], D.B.  Renner\addressmark[MIT], D.
    Richards\addressmark[JLAB], K.  Schilling\addressmark[UWUP], and W.
    Schroers\addressmark[MIT] }

\begin{document}

\begin{abstract}
  The lowest three moments of generalized parton distributions are calculated in full QCD and provide new insight into the behavior of nucleon electromagnetic form factors, the origin of the nucleon spin, and the transverse structure of the nucleon.
  \vspace{1pc}
\end{abstract}
\maketitle
\section{INTRODUCTION}
\label{sec:introduction}
Matrix elements of the light cone operator\\ [.2cm]
{\small
$
 {\cal O}(x) \!=\!\int \!\frac{d \lambda}{4 \pi} e^{i \lambda x} \bar
  \psi (\frac{-\lambda n}{2})\!\!
  \not n {\cal P} e\!^{-ig \int_{-\lambda / 2}^{\lambda / 2} d \alpha \, n
    \cdot A(\alpha n)}\!
  \psi(\frac{\lambda n}{2})
  $ } \\[.2cm]  
  and the tower of twist-two operators 
  $${\cal O}_q^{\lbrace\mu_1\mu_2\dots\mu_n\rbrace} = \overline{\psi}_q
  \gamma^{\lbrace\mu_1} \imag{D}^{\mu_2} \dots
  \imag{D}^{\mu_n\rbrace} \psi_q $$
  provide a wealth of precise information about the quark and gluon structure of the nucleon.   The 
diagonal nucleon matrix element $\langle P |{\cal O}(x) | P \rangle$ 
measures the light cone momentum distribution, $ q(x) $,  and
$ \langle P | {\cal
  O}_q^{\lbrace\mu_1\mu_2\dots\mu_n\rbrace} | P \rangle$ specifies the $(n-1)^{th}$ moment of this distribution, $\int
dx\, x^{n-1} q(x) $. Off-diagonal matrix elements of ${\cal O}(x)$ measure the 
generalized parton distributions\cite{Muller:1998fv}   $ H(x, \xi, t)$ and  $ E(x, \xi, t)$:
$$
\langle P' |{\cal O}(x) | P \rangle = \langle\!\langle \gamma \rangle\!\rangle  H(x, \xi, t) +  \frac{\mbox{i} \Delta} {2 m}  \langle\!\langle \sigma \rangle\!\rangle E(x, \xi, t),
$$
where $\Delta^\mu = P'^\mu - P^\mu$, $ t = \Delta^2$, $\xi = -n \cdot \Delta/2$, and  $\langle \!
\langle \Gamma \rangle \! \rangle = \bar U(P') \Gamma U(P)$. Off-diagonal matrix elements of the  twist-two operators 
$ \langle P' | {\cal
  O}_q^{\lbrace\mu_1\mu_2\dots\mu_n\rbrace} | P \rangle$ yield moments of these generalized parton distributions, and this work considers the generalized form factors 
$A_{n0}(t)  \equiv  \int dx x^{n-1}H(x, 0, t)$ and
$B_{n0}(t) \equiv \int dx x^{n-1} E(x, 0, t)$. 
  
 Two special cases are important for the present work.  The zeroth moments correspond to the familiar electromagnetic form factors (weighted with appropriate quark charges), $A_{10}(t) = F_1(t)$ and $B_{10}(t) = F_2(t)$. The first moments yield the total quark angular momentum, $J_q=\frac{1}{2}[A_{20}(0) + B_{20}(0)]$. Combined with the angular momentum from the quark spin, $\frac{1}{2}\Sigma = \frac{1}{2}[\langle1\rangle_{\Delta u} + \langle1\rangle_{\Delta d}]$, this enables decomposition of the quark contribution to the nucleon spin.

Burkardt\cite{Burkardt:2002hr} has shown that  the generalized parton distribution $H(x, 0, \Delta^2)$ is the Fourier transform of the impact parameter dependent parton distribution
$$
q(x,b_{\perp}) = \int \frac{d^2 \Delta_\perp}{(2 \pi)^2} H(x, 0, -\Delta_\perp^2 )e^{- i b_\perp \Delta_\perp},
$$
where $q(x,b_{\perp})$ is the probability of finding a quark with longitudinal momentum fraction $x$ and transverse position (or impact parameter) $b_\perp$, and $\Delta_\perp$ is the transverse momentum transfer. Physically, we expect 
the transverse size of the nucleon to decrease significantly as $x$ increases. As $ x \to 1$, a single parton carries all the momentum,  the proton wave function is reduced to a single Fock space component, and the transverse wave function has zero spatial extent. 
Since $H(x,0,t)$ is the Fourier transform of the transverse distribution, the slope in $t=\Delta^2_{\perp} $ at the origin measures the rms transverse radius.  The slope should decrease significantly with $x$ and  as $x \to 1$, it should approach zero. Hence, the slope of the generalized form factors $ A_{n0}(t)$ should decrease with increasing $n$, and we demonstrate this effect  in our lattice calculations.

\section{LATTICE CALCULATION}

		\begin{figure}[t]
		 \vspace*{-0.8cm}
		  \includegraphics[clip=false,scale=0.29]{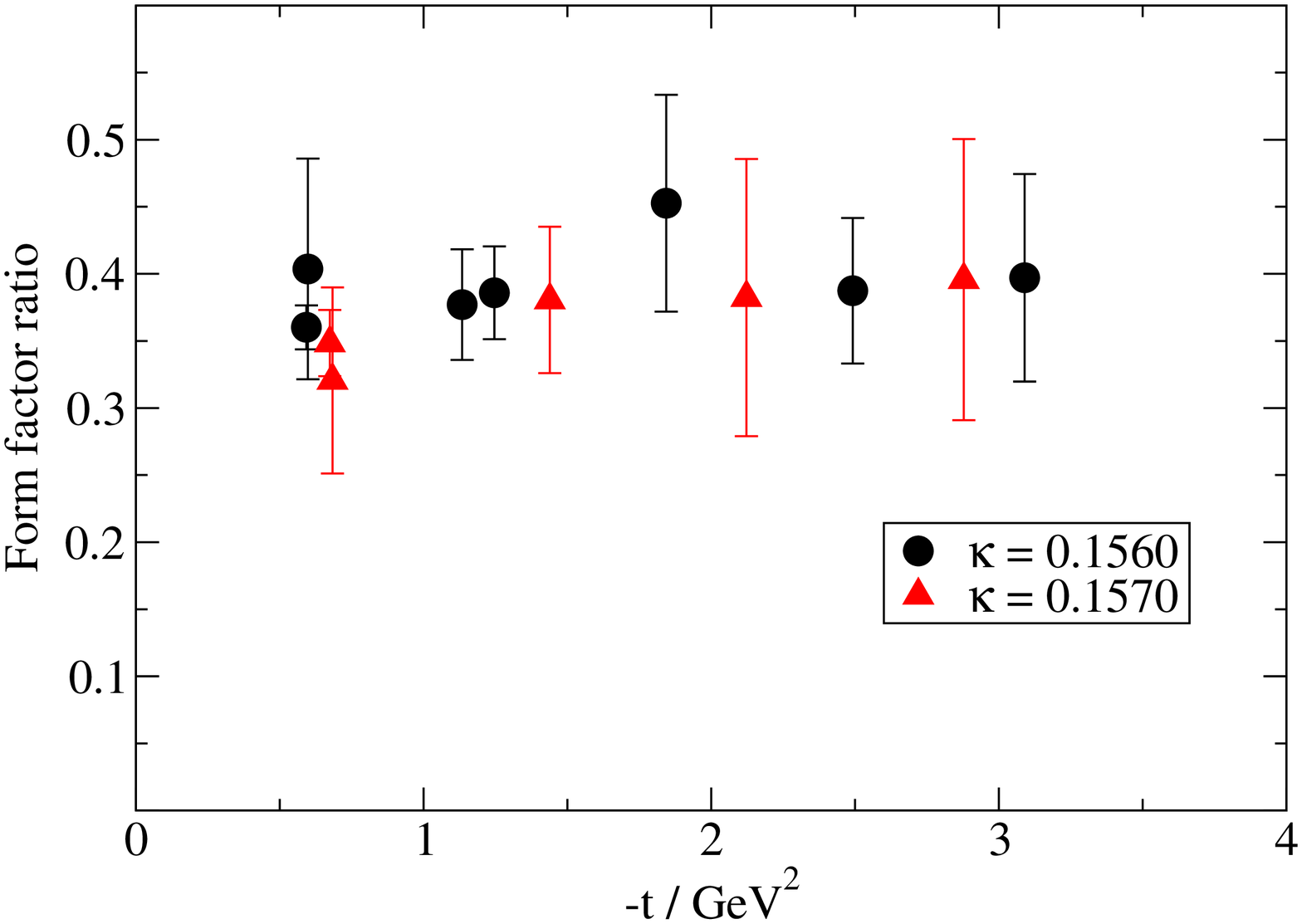}
	  \vspace*{-1.0cm}
		  \caption{Electromagnetic form factor ratio 
		    $\frac{Q^2 F_2(Q^2)}{\log^2 (Q^2 / \Lambda^2) F_1(Q^2)}$.}
		  \label{fig:ji}
		  \vspace*{-.5cm}
		\end{figure}

By calculating an overdetermined set of lattice observables to improve the statistical accuracy of generalized form factors as described in ref.~\cite{Hagler:2003jd}, we calculate
the lowest three moments\\[.3cm]
$  \langle P' | {\cal O}^{\mu_1} | P \rangle \!=\!
  \langle \! \langle \gamma^{\mu_1 }\rangle \! \rangle A_{10}(t) 
  + \frac{\imag}{2 m} \langle \! \langle \sigma^{\mu_1 
    \alpha} \rangle \! \rangle
  \Delta_{\alpha} B_{1
    0}(t), \\ [.3cm]
  \langle P' | {\cal O}^{\lbrace \mu_1 \mu_2\rbrace} | P \rangle =
  \bar P^{\lbrace\mu_1}\langle \! \langle
  \gamma^{\mu_2\rbrace}\rangle  \! \rangle
  A_{20}(t) \\ 
  \mbox{\,\,\,} + \frac{\imag}{2 m} \bar P^{\lbrace\mu_1} \langle \! \langle
  \sigma^{\mu_2\rbrace\alpha}\rangle \! \rangle \Delta_{\alpha} B_{2 0}(t)
  +\frac{1}{m}\Delta^{\{ \mu_1}   \Delta^{ \mu_2 \} }
  C_{2}(t), \\[.3cm]
  \langle P' | {\cal O}^{\lbrace\mu_1 \mu_2 \mu_3\rbrace} | P \rangle
  = \bar P^{\lbrace\mu_1}\bar P^{\mu_2} \langle \! \langle
  \gamma^{\mu_3\rbrace}
  \rangle \! \rangle A_{30}(t)\\ 
   \mbox{\quad}+ \frac{\imag}{2 m} \bar P^{\lbrace \mu_1}\bar P^{\mu_2}
  \langle \! \langle \sigma^{\mu_3\rbrace\alpha} \rangle \! \rangle
  \Delta_{\alpha} B_{3 0}(t) \\
   \mbox{\quad} + \Delta^{\lbrace \mu_1}\Delta^{\mu_2} \langle \! \langle
  \gamma^{\mu_3\rbrace}\rangle \! \rangle A_{32}(t) \\
   \mbox{\quad} + \frac{\imag}{2 m} \Delta^{\lbrace\mu_1}\Delta^{\mu_2}
  \langle \! \langle \sigma^{\mu_3\rbrace\alpha}\rangle \! \rangle
  \Delta_{\alpha} B_{3 2}(t)$,\\[.3cm]
where $\bar P_{\mu} = (P_{\mu} + P'_{\mu})/2 $.

\begin{figure}[t]
 \vspace*{-0.75cm}
     \includegraphics[scale=0.30,clip=true,angle=270]{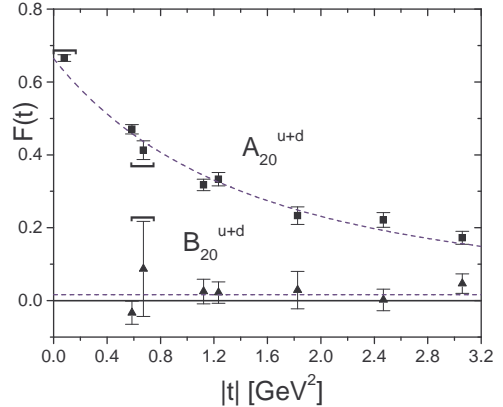}
      \vspace*{-0.8cm}
  \caption{Generalized form factors $A^{\mbox{\tiny
    u+d}}_{20}(t)$ and $B^{\mbox{\tiny
    u+d}}_{20}(t)$, with dipole fits denoted by dashed curves. }
  \label{fig:A2B2}
   \vspace*{-.5cm}
\end{figure}

We calculated connected diagram contributions using  $\sim 200$   SESAM\cite{SESAM} full QCD configurations with Wilson fermions at $\beta = 5.6$ on $16^3 \times 32$ lattices at each of three quark masses, $\kappa$= 0.1570, 0.1565, and 0.1560, corresponding to pion masses defined by $r_0$ of 744, 831, and 897 MeV respectively. 

Figure 1 shows our result for the electromagnetic form factor ratio $F_2 /F_1$ divided by the next to leading order light cone wave function result\cite{Belitsky:2002kj} ${\log^2 (Q^2 / \Lambda^2)} / {Q^2}$ with $\Lambda$ = 0.3 GeV.  The $Q^2$ dependence  is in excellent agreement with the recent JLab data\cite{Gayou:2001qd} plotted in Ref.~\cite{Belitsky:2002kj}, but the overall ratio is a factor of four too high in the heavy quark world in which $m_\pi \sim$ 700-900 MeV.

The total quark contribution to the nucleon spin is given by the extrapolation to $t=0$ of  $A^{\mbox{\tiny u+d}}_{20}(t)$ and $B^{\mbox{\tiny  u+d}}_{20}(t)$ shown in Figure 2.  Since  $A^{\mbox{\tiny u+d}}_{20}(t)$ is calculated directly at   $t=0$  and  $B^{\mbox{\tiny  u+d}}_{20}(t)$ is well fit by a constant that is measured to be nearly zero with small errors, the connected contribution to the angular momentum is measured to within a few percent. Combined with the results of $\Sigma$ from reference \cite{Dolgov:2002zm}, we obtain the connected diagram contributions to the decomposition of nucleon spin shown in Table 1. Similar results have been obtained in refs.~\cite{Gockeler:2003jf,Mathur:1999uf}.      To the extent that the disconnected diagrams do not change the qualitative behavior, we conclude that of the order of 70\% of the spin of the nucleon arises from the quark spin and a negligible fraction arises from the quark orbital angular momentum in a heavy pion world where $m_\pi \sim$ 700 - 900 MeV. 
In the chiral limit, the quark spin contribution must decrease to  $\sim$ 30\% to agree with experiment.

\begin{table}[t]
  \begin{tabular}[b]{*{3}{c|}c}
    \hline
         $\kappa $& 0.1570 & 0.1565 &  0.1560 \\ \hline
          $\Delta \Sigma$ &0.67$\pm$ .04 & 0.73 $\pm$ .03 & 0.68 $\pm$ .02\\ \hline   
          $2 L_q$ &0.06$\pm$ .05 & -0.04 $\pm$ .04 & 0.00 $\pm$ .03\\ \hline
          $2 J_q$ &0.73$\pm$ .04 & 0.69$\pm$ .02 & 0.68 $\pm$ .03\\ \hline
    \end{tabular}
  \caption{Fraction of nucleon spin arising from quark spin, $\Delta \Sigma$, quark orbital angular momentum,  $2 L_q$, and quark total angular momentum,  $2 J_q.$ }
  \label{tab:latt-pars}
\end{table}

Figure 3  shows the generalized form factors  $A^{\mbox{\tiny u+d}}_{10}(t)$,$A^{\mbox{\tiny u+d}}_{20}(t)$, and $A^{\mbox{\tiny u+d}}_{30}(t)$  for $\kappa $ = 0.1560 and 0.1570, corresponding to the lowest three moments of $H(x, 0, t)$.   As explained above, the decrease in slope with increasing moment is a clear manifestation of the decrease of the transverse size of the light cone wave function as $x \to 1$. Note that the error bands are sufficiently narrow that the dramatic change of slope is clearly determined, strongly ruling out a factorized Ansatz for the momentum transfer dependence of generalized form factors. Qualitatively similar behavior is obtained for the connected contributions to $A^{\mbox{\tiny u+d}}_{n0}(t)$ and for the spin-dependent $\tilde{A}^{\mbox{\tiny u-d}}_{n0}(t)$\cite{Schroers}.

\begin{figure}[t]
  \begin{tabular}{*{3}{c}}
    \includegraphics[width=0.47\textwidth,clip=true,angle=0]{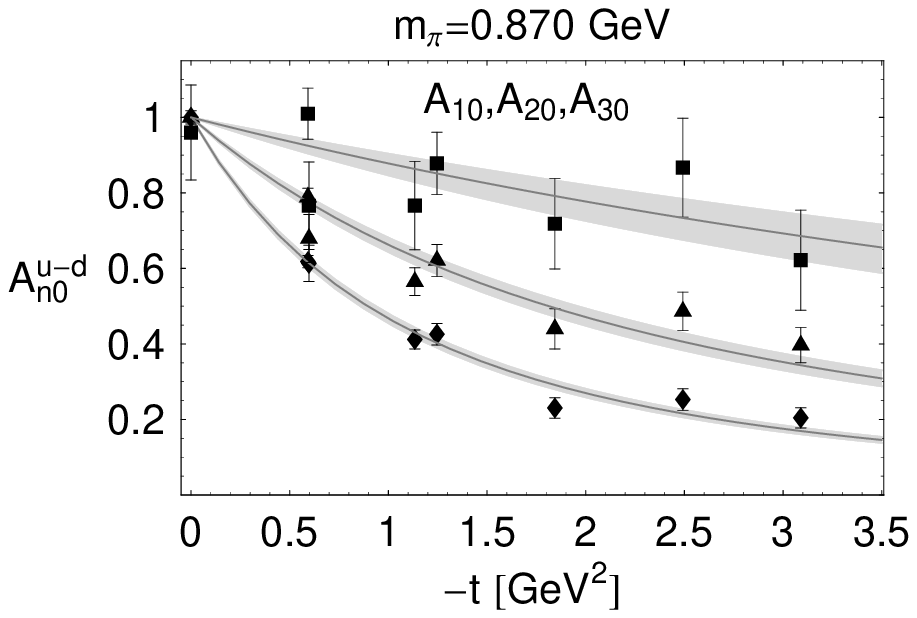}\\
    \includegraphics[width=0.47\textwidth,clip=true,angle=0]{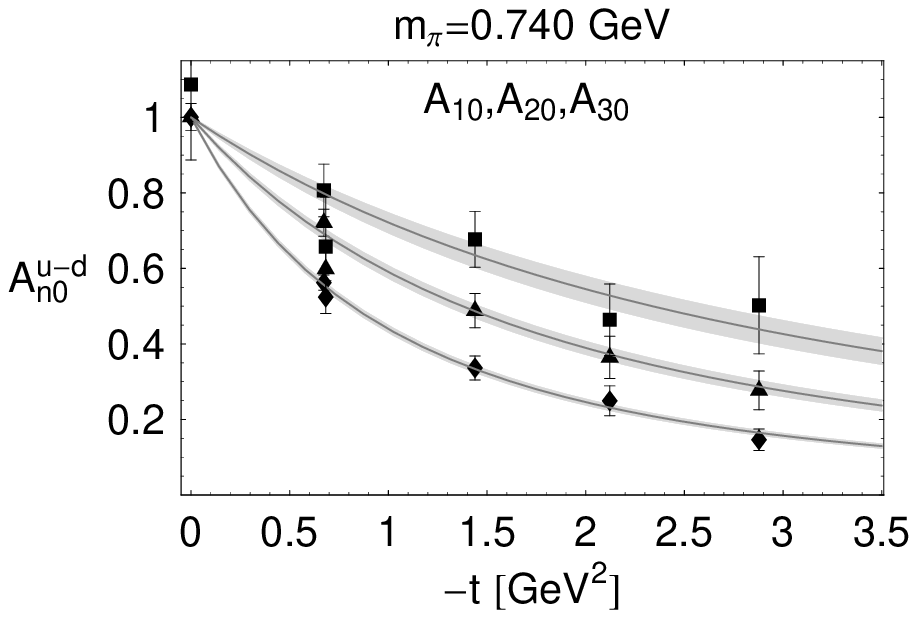}\\
  \end{tabular}
  \vspace*{-0.8cm}
  \caption{\label{fig:slopes} Normalized generalized 
  form factors $A^{\mbox{\tiny u-d}}(t)$ 
  for n=1 (diamonds), n=2 (triangles) and n=3 (squares).  }
\end{figure}


\begin{thebibliography}{99} 


\bibitem{Muller:1998fv}
D.~M{\"u}ller, D.~Robaschik, B.~Geyer, F.~M.~Dittes and J.~Horejsi,
Fortsch.\ Phys.\  {\bf 42} (1994) 101 [arXiv:hep-ph/9812448],
X.~D.~Ji,
Phys.\ Rev.\ Lett.\  {\bf 78} (1997) 610 [arXiv:hep-ph/9603249],
A.~V.~Radyushkin,
Phys.\ Rev.\ D {\bf 56} (1997) 5524 [arXiv:hep-ph/9704207].

\bibitem{Burkardt:2002hr}
M.~Burkardt,
Int.\ J.\ Mod.\ Phys.\ A {\bf 18}, 173 (2003) 
[arXiv:hep-ph/0207047].

\bibitem{Hagler:2003jd}
P.~Hagler, J.~Negele, D.~B.~Renner, W.~Schroers, T.~Lippert and K.~Schilling
                  [LHPC collaboration],
Phys.\ Rev.\ D {\bf 68}, 034505 (2003)
[arXiv:hep-lat/0304018].

\bibitem{SESAM}
N. Eicker et al (SESAM coll.), 
Phys.\ Rev.\ D {\bf 59}, 014509 (1999)
[arXiv:hep-lat/9806027].


	




\bibitem{Belitsky:2002kj}
A.~V.~Belitsky, X.~d.~Ji and F.~Yuan,
Phys.\ Rev.\ Lett.\  {\bf 91}, 092003 (2003)
[arXiv:hep-ph/0212351].


\bibitem{Gayou:2001qd}
O.~Gayou {\it et al.}  [Jefferson Lab Hall A Collaboration],
Phys.\ Rev.\ Lett.\  {\bf 88}, 092301 (2002)
[arXiv:nucl-ex/0111010].

\bibitem{Dolgov:2002zm}
D.~Dolgov {\it et al.}  [LHPC collaboration],
Phys.\ Rev.\ D {\bf 66}, 034506 (2002)
[arXiv:hep-lat/0201021].


\bibitem{Gockeler:2003jf}
M.~G{\"o}ckeler, R.~Horsley, D.~Pleiter, P.~E.~Rakow, A.~Sch{\"a}fer,
G.~Schierholz and W.~Schroers [QCDSF Collaboration] (2003)
[arXiv:hep-ph/0304249].


\bibitem{Mathur:1999uf}
N.~Mathur, S.~J.~Dong, K.~F.~Liu, L.~Mankiewicz and N.~C.~Mukhopadhyay,
Phys.\ Rev.\ D {\bf 62}, 114504 (2000)
[arXiv:hep-ph/9912289].
\bibitem{Schroers}
W. Schroers {\it et al.}, these proceedings, [arXiv:hep-lat/0309065].

%


\end{thebibliography}
\end{document}